\documentclass[12pt]{article}
\usepackage{amssymb}
\usepackage{epsfig}
\usepackage[english]{babel}
\usepackage{float}
\usepackage{slashed}
\newcommand{\be}{\begin{equation}}
\newcommand{\ee}{\end{equation}}
\newcommand{\bea}{\begin{eqnarray}}
\newcommand{\eea}{\end{eqnarray}}

\def\p{\partial}
\def\pslash{\p\raise.3ex \hbox{\kern-.5em /}}
\def\delslash{\nabla\raise.3ex \hbox{\kern-.7em /}}

\begin{document}

\vskip 5cm

\begin{center}
\Large{ \textbf{ Non-Hermitian $\cal PT$ - symmetric relativistic
quantum mechanics with a maximal mass in an external magnetic
field }}
\end{center}
\vskip 0.5cm\begin{center} \Large{V.N.Rodionov}
\end{center}
\vskip 0.5cm
\begin{center}
{Plekhanov Russian University, Moscow, Russia,  \em E-mail
vnrodionov@mtu-net.ru}
\end{center}

\begin{center}

\abstract{Starting with the modified Dirac equations for free
massive particles with the $\gamma_5$-extension of the physical
mass $m\rightarrow m_1 + \gamma_5 m_2$, we consider equations of
relativistic quantum mechanics in the presence of an external
electromagnetic  field. The new approach is developing on the
basis of  existing methods for study the unbroken $\cal PT$
symmetry of  Non-Hermitian Hamiltonians. The paper shows that this
modified model contains the definition of the mass parameter,
which may use as the determination of the magnitude scaling of
energy $M$. Obviously that the transition to the standard approach
is valid when small in comparison with $M$ energies and momenta.
Formally, this limit is performed when $M\rightarrow\infty$, which
simultaneously should correspond to the transition to a
\emph{Hermitian limit: }$m_2\rightarrow 0$. Inequality $m \leq M$
may be considered and as the restriction of the mass spectrum of
fermions considered in the model. Within of this approach, the
effects of possible \emph{observability} mass parameters: $m_1,
m_2$, $M$ are investigated taking into account the interaction of
the magnetic field with charged fermions together with the
accounting of their anomalous magnetic moments.}


\end{center}


\section{Introduction}

As it is well-known the idea about existence of a maximal mass in
a broad spectrum mass of elementary particles at the  Planck mass
was suggested by Moisey Markov in 1965 \cite{Markov} \be
\label{Markov} m\leq m_{Planck}\cong 10^{19} GeV. \ee The
particles with the limiting mass
$$
                 m = m_{Planck},
$$
named by the author "maximons" should play a special role in the
world of elementary particles. However, Markov's original
condition (\ref{Markov}) was purely phenomenological and he used
standard field theoretical techniques even for describing the
maximon.
  In the end of the seventies of the last century
    a more radical approach was developed by V.G.
Kadyshevsky \cite{Kad1}.   His model  contained a limiting mass
$\mathfrak{M}$ as \emph{a new fundamental physical constant}.
Doing this condition of finiteness  of the mass spectrum should be
introduced by the relation: \be\label{Mfund1} m \leq
\mathfrak{M},\ee where a new constant $\mathfrak{M}$ was named by
the \emph{fundamental mass}.

 Really in the papers
\cite{Kad1}-\cite{Rod} the existence of mass $\mathfrak{M}$ has
been understood as a new principle of Nature similar to the
relativistic and quantum postulates, which was put into the ground
of the new quantum field  theory. At the same time the new
constant $\mathfrak{M}$ is introduced in a purely geometric way,
like the velocity of light is the maximal velocity in the special
relativity.

Indeed, if one chooses a geometrical formulation of the quantum
field theory (QFT), the adequate realization of the limiting mass
hypothesis  is reduced to the choice of the de Sitter geometry as
the geometry of the 4-momentum  space of the constant curvature
with a radius equal to $\mathfrak{M}$ \cite{Kad1}.

Besides the de Sitter space there is another  space of constant
curvature, breaking into a Minkowski space in small 4-momentum,
which is called the space of anti- de Sitter. The detailed
analysis of the different aspects of the construction of the
modified quantum field theory with the maximal mass in the curved
momentum anti-de Sitter space has allowed to obtain a number of
interesting results. In particular, it has been shown that
non-Hermitian fermionic Hamiltonians with the $\gamma_5$-dependent
mass term must arise in the modified field theory (see, for
example,\cite{Max},\cite{KMRS} ).

Now it is well-known fact, that the reality of the spectrum in
models with a non-Hermitian Hamiltonian is a consequence of $\cal
PT$ -invariance of the theory, i.e. a combination of spatial and
temporary parity of the total Hamiltonian: $[H,{\cal PT}]\psi =0$.
When the $\cal PT$ symmetry is unbroken, the spectrum of the
quantum theory is real. These results explain the growing interest
in this problem. For the past a few years has been studied a lot
of new non-Hermitian $\cal PT$-invariant systems \cite{Rod1} -
\cite{neznamov2}. It is important to note that the previous works
which were devoted to studying pseudo-Hermitian quantum mechanics
  with $\gamma_5$-mass contribution \cite{ft12}-\cite{ft13}  the
restrictions of mass parameters  were presented only as
${m_1}^2\geq {m_2}^2$,$\,m\leq m_1.$

This paper has the following structure. In section II the
non-Hermitian approach to the construction of plane wave solutions
is formulated for the case  free massive particles. In the third
section we study the basic characteristics of modified Dirac
models with $\gamma_5$-massive contributions in the external
magnetic field. Then, in the fourth section, we consider the
modified  Dirac-Schwinger-Pauli model  in the magnetic field with
non-Hermitian extension.  This section also contains the
discussion of  the effects of the possible observability of the
parameters: $m_1, m_2$ and $M$ taking into account the interaction
with the magnetic field of charged fermions together with regard
to their anomalous magnetic moments. The fifth section contains
summary and conclusion.

\section{Non-Hermitian extensions of plane waves}

 Let us now consider the solutions of  modified Dirac equations for free
massive particles  following from procedures of the
$\gamma_5$-extension of the mass $ m\rightarrow m_1 +\gamma_5
m_5$:

\be\label{D2} \Big(i\partial_\mu\gamma^{\mu}-m_1-\gamma_5 m_2
\Big) \widetilde{\psi}(x,t)=0. \ee It is obvious that the
Hamiltonian associated with this equation is non-Hermitian due to
the appearance in it the $\gamma_5$-dependent mass components
($H\neq H^{+}$).

First-order equations (\ref{D2}) can be transformed into equations
of the second order by applying to (\ref{D2}) the operator:
\be\label{D22} \Pi=i\partial_\mu\gamma^{\mu}+ m_1-\gamma_5 m_2
.\ee In a result the modified Dirac equation converts  to the
Klein-Gordon equation: \be \label{KG}
\left(\partial^2+m^2\right)\psi(x,t)=0 \label{e20} \ee where the
physical mass of particle $m$ is expressed through the parameters
$m_1$ and $m_2$ \be \label{012} m^2={m_1}^2- {m_2}^2. \ee It is
easy to see from (\ref{012}) that the  mass $m$, appearing in the
equation (\ref{KG}) is real, when the inequality \be \label{e210}
{m_1}^2\geq {m_2}^2.\ee is accomplished.

A.Mustafazadeh identified the necessary and sufficient
requirements of reality  of eigenvalues for pseudo-Hermitian and
$\cal PT$-symmetric Hamiltonians and formalized the use these
Hamilton operates  in his papers \cite{alir},\cite{ali} and \cite
{spec}-\cite{most5}. According to the  recommendations of this
works we can define Hermitian operator $\eta$, which transform
non-Hermitian Hamiltonian
 by means of invertible
transformation to the Hermitian-conjugated one. It is easy to see
that with Hermitian operator \be \label{et1} \eta= e^{\gamma_5
\vartheta} \ee we can obtain
 \be\label{D3}  \eta H \eta^{-1}= H^{+},\ee
 where
  \be\label{H} H =\alpha p+ \beta(m_1
+\gamma_5 m_2)\ee  and \be\label{H+} H^+ =\alpha p+ \beta(m_1
-\gamma_5 m_2).\ee Here  matrices
$\alpha_i=\gamma_0\cdot\gamma_i$, $\beta=\gamma_0$, and
$\vartheta= \textmd{arctanh} (m_2/m_1)$

In addition, multiplying the Hamilton operator $H$ from left to $
e^{{\vartheta\gamma_5 }/2}$ and taking into account that matrices
$\gamma_5$ commute with matrices $\alpha_i$ and anti-commute with
$\beta$, we can obtain
 \be \label{H0}  e^{{\gamma_5 \vartheta}/2}
H = H_0  e^{{\gamma_5 \vartheta}/2},\ee where $H_0 =\alpha p
+\beta m $ is a ordinary Hermitian Hamiltonian of a free particle.

The mathematical sense of the action of the operator (\ref{et1})
it turns out, if we notice that according to  the properties of
$\gamma_5$ matrices, all the even  degree of $\gamma_5$ are equal
to 1, and all odd degree are equal to $\gamma_5$. Given that
$\cosh(x)$ decomposes on even and $\sinh(x)$ odd degrees of $x$,
the expressions (\ref{D3})-(\ref{H0}) can be obtained by
representing non-unitary exponential operator $\eta$ in the form
\be\label{eta} \eta=e^{\gamma_5 \vartheta}=\cosh\vartheta+\gamma_5
\sinh\vartheta,\ee  where \be\label{chsh} \cosh\vartheta =
m_1/m;\,\,\,\, \sinh\vartheta = m_2/m. \ee

 The region of the unbroken
$\cal PT$-symmetry of (\ref{H}) can  be found  in the form
(\ref{e210}). However, it is not apparent that the area with
undisturbed $\cal PT$-symmetry defined by such a way does not
include  the regions, corresponding to the  some unusual
particles, description of which radically distinguish from
traditional one.

A feature of the model with $ \gamma_5 $-mass contribution is that
it may contain any additional restrictions for mass parameters
besides (\ref{e210}). Indeed while that for the physical mass $m$
one may be constructed by infinite number combinations of $ m_1 $
and $ m_2 $, satisfying to (\ref{012}), however besides it need to
provide and the rules of conformity of this parameters in the
Hermitian limit.

In particular, the simple a linear scheme  may be easy constructed
if one takes the obvious restriction for the mass spectrum of  the
fermions  in the form \be \label{m1111} m \leq M1, \ee where $M1 $
- the fixed value of mass parameter $ m_1$($M_1=m_1$). Using this
approach we can in principle describe   the whole spectrum of
fermions when $m \leq M_1$ by means of defining appropriate values
$ m_2 $. According to (\ref{012}) one can obtain the expression
\be \label{M11}  m = M_1\sqrt { 1 - {m_2} ^ 2/{M_1}^2 }. \ee  With
the help (\ref {M11}) we can see,  that when the mass $ m_2 $ is
increased,  the values of the physical mass tends  to zero.  The
equality of parameters $ m_2 = m_1 $ corresponds to the case of
massless fermions. But it should be noted that in this model the
Hermitian limit $ m_2 \rightarrow 0 $ may be  reached only in the
case of the the particles with maximal mass $ m = M1 $. At the
same time,  \emph{the Hermitian limit is absent for  all other
 mass values}.

 Therefore the procedure of limitations of  the
physical mass spectrum by the inequality $ m \leq M1 $ has the
essential drawback since in this frame is not possible to describe
all ordinary fermions, respecting the  Principle of Conformity,
except, $ m = M1 $.

These considerations make search in the frame of mass restriction
of $m\leq m_1$ the existence of more complicated non-linear
dependence of limiting mass value \be\label{mM} m \leq
M(m_1,m_2),\ee which meets the requirements of the  Principle of
Conformity:

i)The Dirac limit must exist for all \emph{ordinary fermions} for
which
 the condition(\ref{mM}) is satisfied.

ii)In Hermitian limit  the parameter $m_1$ must coincide with the
physical mass $m$.

 The fulfillment of these conditions lets to find the most
appropriate scheme  restriction of  the mass fermions,  for  which
exist  the consistency with the ordinary Dirac theory.

Possible explicit expression for $M(m_1,m_2)$, may be obtained
from the simple mathematical theorem about the arithmetical
average of two non-negative real numbers $a$ and $b$ which is
always  greater than or equal to the geometrical mean of the same
numbers.
 Really, let $a=m^2$ and $b={m_2}^2$ then using
$$ \frac{{m}^2+{m_2}^2}{2}\geq\sqrt{m^2\cdot {m_2}^2}$$ and
substitution (\ref{012}), we can get the inequality \be\label{mM1}
m\leq {m_1}^2/2m_2 = M(m_1,m_2).\ee Values  $M$ now is defined by
two parameters $m_1,m_2$ and in the limit $m_2 \rightarrow 0$,
value of the maximal mass $M$ tends to infinity. It is very
important that in this limit  the restriction of mass value of
particles  completely disappear. In such a way  a natural
transition   of the modified model to  the ordinary Dirac theory
is demonstrated for any values of the physical mass.

Using (\ref{012}) and expression (\ref{mM1}) we can also obtain
the system of two equations \be\label{sys}\left\{
\begin{array}{c}
  m={m_1}^2-{m_2}^2 \bigskip\\
  M={{m_1}^2}/{2 m_2} \\
\end{array}\right.
\ee The solution of this system  relative to the parameters $m_1$
and $m_2$ may be represented in the form

\be\label{m11111} {m_1}^{\mp} =\sqrt{2}M
\sqrt{1\mp\sqrt{1-m^2/{M}^2}}, \ee

\be \label{m22222}{m_2}^\mp =  M\left(1\mp \sqrt{1-m^2/{M}^2}
\right). \ee

It is easy to verify that the obtained values of the mass
parameters satisfy the conditions (\ref{012}) and (\ref{e210})
regardless of which sign is choosen. Besides it should be
emphasized that,  formulas (\ref{m11111}),(\ref{m22222}) in the
case of the upper sign  are agreed with conditions $m_2\rightarrow
0$ and $m_1\rightarrow m$ when $M \rightarrow \infty,$ i.e. there
are a so-called Hermitian and Dirac limits, which determined in
the conditions i) and ii). However, if one choose a lower sign
(i.e. for the ${m_1}^+$ and ${m_2}^+$) the such limits are absent.
Thus we can see that the nonlinear scheme of mass restrictions
(see \ref{mM1}) additionally contains the solutions to satisfy the
requirements of the "exotic" particles. However this "exotic"
solutions should be considered  only as an indication  of the
principal possibility of the existence of such particles. In this
case, as follows from (\ref{m11111}),(\ref{m22222}) for each
ordinary particles may be exist some "exotic" partners, possessing
the same mass, but a number of unique properties. \footnote{As the
exotic particles do not agree in the limit $m_2 \rightarrow 0$
with the ordinary Dirac expressions then one can assume that in
this case we deal with a description of some new particles,
properties of which have not yet been studied. This fact for the
first time has been fixed by V.G.Kadyshevsky in his early works in
the geometric approach to the development of the theory with a
"fundamental mass" \cite{Kad1}. Besides in \cite{Max},\cite{KMRS}
it was noted that the most intriguing prediction of the new
approach is the possible existence of exotic fermions with no
analogues in the SM, which may be candidates for dark matter
constituents.}

\begin{figure}[h]
\vspace{-0.2cm} \centering
\includegraphics[angle=0, scale=0.5]{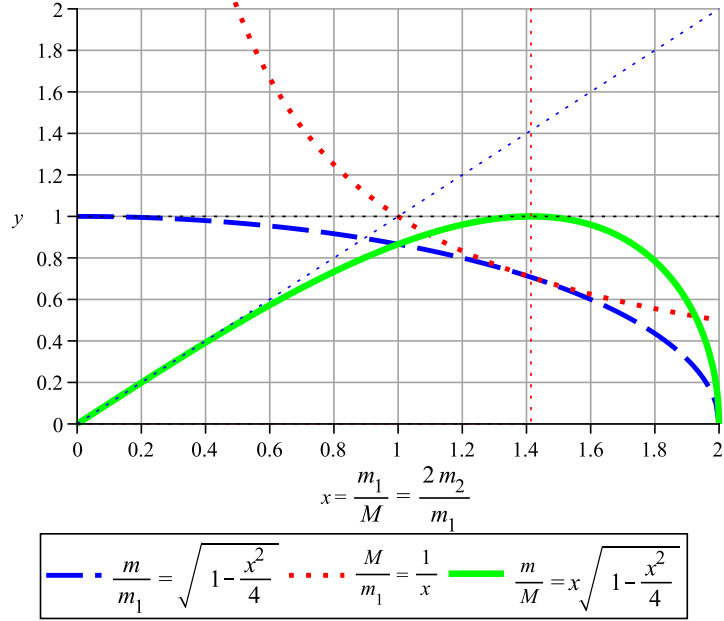}
\caption{Dependence of $m/M,  M/m_1$, and  $m/m_1 $ from the
parameter $x=m_1/M=2m_2/m_1$} \vspace{-0.1cm}\label{Fig.1-1}
\end{figure}

Let's consider the  "normalized" parameter of the modified model
with the maximal mass $M$: \be \label{x} x=\frac{m_1}{M}=\frac{2
m_2}{m_1}\ee and using (\ref{M11}) we can obtain \be\label{y}
\frac{m_2}{M}=\frac{x^2}{2},\ee \be
\frac{m}{M}=x\sqrt{1-x^2/4}.\ee

At Fig.1 we can see dependence of the the normalized parameters
$m/m_1$, $M/m_1$ and $m/M$ on  the parameter $x=m_1/M=2m_2/m_1$.
In particular, maximum value of the particle  mass $ m/M = 1 $ is
achieved at the value of $ x = \sqrt{2} $. In this point parameter
$m_2/M$ is also equal to unit. Further increasing of $x$, leads to
decreasing of $m/M$ and at the point $x=2$  this value is equal to
zero. Thus, the region of $x > \sqrt{2}$ ($m_2/M > 1$ )
corresponds to the description of the "exotic particles", for
which  there is not  transition to Hermitian limit.

 In the frame of the  inequality  (\ref{e210})  we can see
 three specific sectors of unbroken $\cal PT$-symmetry of the
 Hamiltonian (\ref{H}) in the plane
$\nu_1=m_1/M, \nu_2=m_2/M$.  Thus the plane $\nu_1,\nu_2$ may be
divided by the three groups of the inequalities:

$$I. \,\,\,\,\,\,\,\,\,\,\,\,\,\,\,\,\,\,\, \nu_1/\sqrt{2} \leq
\nu_2 \leq\nu_1,$$
 $$II. \,\,\,\,-\nu_1/\sqrt{2}< \nu_2 <
\nu_1/\sqrt{2},$$
 \be\label{ogr}III. \,\,\,\,\,\,\,\, -\nu_1
\leq \nu_2\leq-\nu_1/\sqrt{2}.\ee

Only the area $II.$ corresponds to the description of ordinary
particles, while the $I.$ and$. III.$ agree with the description
of some as yet unknown particles. This conclusion is not trivial,
because in contrast to the geometric approach, where the emergence
of new unusual properties of particles associated with the
presence in the theory a new degree of freedom (sign of the fifth
component of the momentum $\varepsilon=p_5/|p_5|$ \cite{KMRS}), in
the case of a simple extension of the free Dirac equation due to
the additional $\gamma_5$-mass term, the satisfactory explanation
of this fact is not there yet.


Then we can  establish the limits of change of parameters.  At
preset values of $m$ and $M$ as it follows from the
(\ref{m11111}),(\ref{m22222})  the limits of variation of
parameters $m_1$ and $m_2$ are the following: \be\label{limit}
m\leq m_1\leq 2 M; -2M \leq m_2 \leq 2M.\ee  In the areas of
change of these parameters has a point in which
 we have
\be\label{m111}  m_1=\sqrt{2}M;\,\, m_2=M.\ee In this point the
physical mass $m$  reaches its maximum value $m=M$ and corresponds
to mass of  the maximon.

If we used the standard representation of $\gamma$-matrixes the
non-Hermitian Hamiltonian $H$ can be whiten in the following
matrix form
$$H=\left(%
\begin{array}{cccc}
  m_1 & 0 & p_3-m_2 & p_1-ip_2 \\
  0 & m_1 & p_1+ip_2 & -m_2-p_3 \\
  m_2+p_3 & p_1-ip_2 & -m_1 & 0 \\
  p_1+ip_2 & m_2-p_3 & 0 & -m_1 \\
\end{array}%
\right),$$ where $p_i$ are components of momentum.

 It is clear that $$H\psi=E\psi.$$ The condition $\det{(H-E)}=(-E^2 + {m_1}^2-{m_2}^2 + {p_{\bot}}^2 +{p_3}^2)^2 =0$
results in the eigenvalues of $E$ which are represented in the
form: \be\label{E} \emph{E}=\pm\sqrt{{m_1}^2-{m_2}^2 +
{p_{\bot}}^2 +{p_3}^2},\ee where $p_{\bot}=\sqrt{{p_1}^2+{p_2}^2}$
and ${m_1}^2-{m_2}^2 =m^2 $, that coincide with the eigenvalues of
energy  of Hermitian operator $H_0$.

Let us now consider the state of a free particle with certain
values of the momentum and energy, which is described by a plane
wave and can be written as \be\label{psi} \widetilde{\psi}=
\frac{1}{\sqrt{2E}}\widetilde{u}e^{-ipx}. \ee It is easy to see
that the wave amplitude $\widetilde{u} $ is determined by
bispinor, normalization of which now needs an additional
explanation.

Really using (\ref{D2}) and taking into account properties of
matrices $\gamma_o, \vec{\gamma}, \gamma_5$, we can  write also
complex-conjugate equation from equation (\ref{D2})
\be\label{conj} \left(-p_0\tilde{\gamma_0}
-\textbf{p}\widetilde{\vec{\gamma}}-m_1-\gamma_5
m_2\right)\widetilde{\psi^{*}}=0, \ee where $\tilde{\gamma_\mu }$
are transpose matrix.
 Rearranging function $\widetilde{\psi^{*}}$
and introducing new bispinor $\overline{\widetilde{\psi}}=
\widetilde{\psi^{*}}\gamma_0$, we can obtain \be\label{bar}
\overline{\widetilde{\psi}}\left(\gamma p+m_1 -\gamma_5
m_2\right)=0.\ee The operator $p$ is assumed here acts on the
 function, standing on the left of it.
Using (\ref{et1}) we can write equation (\ref{D2}), (\ref{bar}) in
the following form \be \label{1} \left(p\gamma - m
\eta\right)\widetilde{\psi}=0\ee \be\label{2}
\overline{\widetilde{\psi}} \left (p\gamma + m \eta^{-1}
 \right)=0\ee

 Multiplying (\ref{1}) on the left of the
 $\bar{\widetilde{\psi}}
e^{-\vartheta\gamma_5} $ and the equation (\ref{2}) on the right
of the $e^{\vartheta\gamma_5}\widetilde{\psi}$ and summing up the
resulting expressions, one can obtain \be
\bar{\widetilde{\psi}}e^{-\vartheta\gamma_5/2}\gamma_{\mu}
e^{\vartheta\gamma_5/2}(p \widetilde{\psi}) +
(p\widetilde{\bar{\psi}})e^{-\vartheta\gamma_5
/2}\gamma_{\mu}e^{\vartheta\gamma_5/2}=p_\mu\left(\bar{\widetilde{\psi}}e^{-\vartheta\gamma_5
/2}\gamma_{\mu}e^{\vartheta\gamma_5/2}\widetilde{\psi}\right)=0
\ee Here brackets indicate  which of the  function are  subjected
to  the action of  the operator $p$. The obtained equation has the
 form  of the continuity equation

 \be \label{j}\partial_\mu j_\mu =0,
 \ee
 where \be j_\mu = \widetilde{\bar{\psi}}e^{-\vartheta\gamma_5
/2}\gamma_\mu
 e^{\vartheta\gamma_5}\gamma_\mu\widetilde{\psi} = \left(\widetilde{\psi^{*}} e^{\vartheta\gamma_5}\widetilde{\psi}, \widetilde{\psi^{*}}\gamma_0\vec{\gamma }
 e^{\vartheta\gamma_5}\widetilde{\psi} \right)\ee

  Thus here the value of $j_\mu $ is a 4-vector of current density of
  particles in the model with $\gamma_5$-mass extension.
   It is very important that its temporal component \be\label{j_0} j_0=\widetilde{\psi}^{*}e^{\vartheta\gamma_5}\widetilde{\psi} \ee
     does not change in time (see (\ref{j}) and positively defined.
   It is easy to see from the following procedure.
Let us construct  the  norm of any state for considered model for
arbitrary vector, taking into account the weight operator $\eta$
(\ref{eta}):
$$
\widetilde{\psi}= \left( \begin{array}{cc}
 x+i y&{} \\
u+iv&{}\\
z+iw&{}\\
t+ip&{}
\end{array}\right).
$$
Using (\ref{chsh}), (\ref{j_0}),  in a result we have

$$
 \widetilde{\psi^{*}}\eta=\left(
\frac{m_1+m_2}{m}(x-iy),\frac{m_1+m_2}{m}(u-iv),
\frac{m_1-m_2}{m}(z-iw), \frac{m_1-m_2}{m}(t-ip)\right).
$$
Then \be \label{Psi} \langle{}
\widetilde{\psi^{*}}\eta|\widetilde{\psi}\rangle=\frac{m_1+m_2}{m}(x^2+y^2)+\frac{m_1+m_2}{m}(u^2+v^2),
\frac{m_1-m_2}{m}(z^2+w^2),\frac{m_1-m_2}{m}(t^2+p^2) \ee is
explicitly non negative, because  $m_1\geq m_2$ in the area of
unbroken ${\cal PT}$-symmetry (\ref{e210}).

With the help of (3),(7) and properties commutation of
$\gamma$-matrix one can obtain that components of new bispinor
amplitudes which must satisfy the following system of algebraic
equations:

\be\label{u} \left(\gamma{p}-m
e^{\gamma_5\vartheta}\right)\widetilde{u}=0; \ee

\be\label{u1} \overline{\widetilde{u}} \left(\gamma{p}-m
e^{-\vartheta\gamma_5}\right)=0, \ee where
$\overline{\widetilde{u}}={\widetilde{u}}^{*}\gamma_0$.

 According  to (\ref{H0}),(\ref{et1}) one can  write bispinor amplitudes
 in the form

\be\label{u1} \widetilde{u}=\sqrt{2m}\left(%
\begin{array}{c}
  A_1 w \\
  A_2 w \\
\end{array}%
\right); \ee

\be \label{u2}\overline{\widetilde{u}}=\sqrt{2m}\left(%
\begin{array}{cc}
 A_1 w^{*}, & - A_2 w^{*} \\
\end{array}%
\right),\ee where the notations are used:
$$
A_1=\cosh\frac{\vartheta}{2}\cosh\frac{\beta}{2}
+\sinh\frac{\vartheta}{2}\sinh\frac{\beta}{2}(\textbf{n} {\sigma})
;
$$
$$
A_2= \sinh\frac{\vartheta}{2}\cosh\frac{\beta}{2}
+\cosh\frac{\vartheta}{2}\sinh\frac{\beta}{2}(\textbf{n}
{\sigma}).
$$
In addition, we have  relations (\ref{eta}),(\ref{chsh}) and
 $\cosh\beta=\emph{\emph{\emph{E}}}/m,
\sinh\beta=p/m$. And also
   $w $ - two-component spinor, satisfying the normalization condition
$$
   w^{*} w =1.
$$
Besides need to note that $\sigma$ are  $2\times2$-Pauli matrices
and $\textbf{n}=\textbf{p}/p $ - a unit vector in the direction of
the momentum.

The explicit form of these spinors can be found using the
condition that spiral states correspond to the plane wave in which
spinors $w$ is a eigenfunctions of the operator $(\sigma {\textbf
n})$
$$
           \sigma{\textbf n}w^{\zeta}=\zeta w^{\zeta}.
$$
Therefore we get
$$
   w^{1} = \left(%
\begin{array}{c}
  e^{-i\varphi/2}\cos\theta/2 \\
  e^{i\varphi/2}\sin\theta/2 \\
\end{array}%
\right),\,\,\,\,\,                       w^{-1} = \left(%
\begin{array}{c}
  -e^{-i\varphi/2}\sin\theta/2 \\
  e^{i\varphi/2}\cos\theta/2 \\
\end{array}%
\right),
$$
where $\theta$ and $\varphi$ - polar and azimuthal angles that
determine the direction \textbf{n} concerning  to the axes
$x_1,x_2,x_3$.

It is easy to verify by the direct multiplication that really
$$
\overline{\widetilde{u}}\widetilde{u} = 2 m.
$$
This result however in advance obvious, because there are the
connection between bispinor amplitudes of modified equations
$\overline{\widetilde{u}},\widetilde{u}$ and  corresponding
solutions of the ordinary Dirac equations:
 $$ \widetilde{u}= e^{-
\gamma_5\vartheta/2}u$$ $$ \overline{\widetilde{u}}= \overline{u}
e^{\gamma_5\vartheta/2}.
 $$
 Taking into account that  the  Dirac bispinor amplitudes as usually \cite{TKR}
are normalized by invariant condition $\overline{u} u =2m$. Hence
we have \be\label{u1u} \overline{\widetilde{u}}\widetilde{u}
=\overline{u} u = 2m. \ee

By using (\ref{u1u}) we can also obtain
$$
\overline{\widetilde{u}}e^{-\gamma_5 \vartheta}\gamma_\mu
\widetilde{u}=2 p_\mu.
$$
Taking into account (\ref{psi}) one easily finds

$$
    \widetilde{\psi}^{*}j_\mu e^{\gamma_5 \vartheta} \widetilde{\psi}= \{ 1, \textbf{p}/E  \},
$$
whence it follows that the operator $\eta$ in full compliance with
Mostafazadeh's  result (see, for example, \cite{ali},\cite{alir})
induces the inner product
  $$
      \widetilde{\psi}^{*}\eta  \widetilde{\psi}= 1
  $$
for $\widetilde{\psi}\neq 0$.

\section{Dirac modified models with $\gamma_5$-massive
contributions in the external homogenies magnetic field}

As it is known  wave Dirac equations provide a basis for
relativistic quantum
 mechanics and quantum electrodynamics of spinor particles in
 external electromagnetic fields. Solutions of relativistic wave
 equation are referred to as one-particle wave functions which
 allow the development of the approach known as the Furry picture.
This method incorporates study the interactions with the external
field exactly, regardless of the field  intensity \cite{TKR}.
Beside there is no regular methods of describing such an
arbitrariness explicitly. The physically most important exact
solutions of the Dirac equations are: an electron in a Coulomb
field, in a uniform magnetic field and in  the field of a plane
wave. In this connection, is of interest from investigating a
modified non-Hermitian Dirac models which describe an alternative
formulation of relativistic quantum mechanics where the Furry
picture may be realized too.

Consider a uniform magnetic field $\textbf{H}=(0,0,H)$ directed
along the $x_3$ axis ($H > 0$). The electromagnetic potentials are
chosen in the  gage \cite{TKR} \be
   A_0=0,\,\,  A_3=0,\,\, A_1=0,\,\,A_2=H x_1.
\ee We can write the modified Dirac equations in the form
\be\label{cal P}
  \left( \gamma_\mu {\cal P^\mu} -m e^{\vartheta
  \gamma_5}\right)\Psi=0,
\ee were ${\cal P^\mu} = i\partial_{\mu} -e A_\mu $ ; $e=-|e|$ and
$\gamma$- matrixes still chosen in the standard representation. In
the field under consideration, the operators ${\cal P}_0,\,{\cal
P}_2 $ and ${\cal P}_3$ are mutually commuting integrals  of
motion $[{\cal D},{\cal P}_0]=0$,$[{\cal D},{\cal P}_2]=0$,
$[{\cal D},{\cal P}_3]=0$, where ${\cal D}=( \gamma_\mu {\cal
P}^{\mu} - m e^{\vartheta
  \gamma_5}) $.

Let present the function $\Psi $ in the form
$$
         \Psi = \left(%
\begin{array}{c}
  \psi_1 \\
  \psi_2 \\
  \psi_3 \\
  \psi_4 \\
\end{array}%
\right)e^{-i E t}
$$
and use Hamilton's form of Dirac  equations \be \label{H2}H\psi =
E\psi \ee where
$$
H=(\alpha \textbf{{\cal P}}) +\beta m_1 +\beta\gamma_5 m_2.
$$
It is useful to introduce the change of variables \cite{TKR}
$$
  \psi(x_1,x_2,x_3)= e^{ip_2 x_2+ip_3 x_3}\Phi(x_1)
$$
 we can obtain the following system of equations:
\be\label{sist1}(E\mp m_1)\Phi_{1,3}+iR_2 \Phi_{4,2} -(p_3 \mp
m_2) \Phi_{3,1} =0;\ee where $ R_2=\left[\frac{\partial}{\partial
x_1} +(p_2+{e H})\right];$ \be\label{sist}
 (E\mp
m_1)\Phi_{2,4}+iR_1\Phi_{3,1}+(p_3 \pm m_2) \Phi_{4,2}=0; \ee
where $R_1=\left[\frac{\partial}{\partial x_1} -(p_2+{e
H})\right],$ and top mark  relates to the components of the wave
function with the first indexes, and the lower - to the components
with the second indexes.

Next convenient to go to the dimensionless variable \be\rho =
\sqrt{\gamma}x_1 +p_2/\sqrt{\gamma},  \ee where $\gamma=|e| H$,
and equations (\ref{sist1}),(\ref{sist}) take the form
\be\label{PH1}
 (E\mp m_1)\Phi_{1,3} + i
 \sqrt{\gamma}\left(\frac{d}{d\rho}+\rho\right)\Phi_{4,2}-(p_3 \mp
 m_2)\Phi_{3,1} =0;
\ee \be\label{PH2} (E\mp m_1)\Phi_{2,4} + i
 \sqrt{\gamma}\left(\frac{d}{d\rho}-\rho\right)\Phi_{3,1}+(p_3 \pm
 m_2)\Phi_{4,2} =0.
\ee General solution of this system can be represented in the form
of the Hermite functions
$$
u_n(\rho)=\left(\frac{\gamma^{1/2}}{2^n n! \pi^{1/2}}
\right)e^{-\rho^2 /2}H_n(\rho),
$$
where $H_n(x)$ is standardizing the Hermite polynomials:
$$
  {\mathit{H}}_{n}(x)=(-1)^n
e^{x^2/2}\frac{d^n}{dx^n}e^{-x^2/2}.
$$

In should be noted that Hermite function are satisfied to the
recurrence relation: \be\label{u11}
\left(\frac{d}{d\rho}+\rho\right)u_n=(2n)^{1/2}u_{n-1}; \ee
\be\label{u22}
\left(\frac{d}{d\rho}-\rho\right)u_{n-1}=-(2n)^{1/2}u_{n}.\ee It
is easy to see  from (\ref{u11}),(\ref{u22}) that
$$
   \left(\frac{d}{d\rho}+\rho \right)\left(\frac{d}{d\rho}-\rho
   \right)u_n = -2n u_n
$$
and hence ( see, for example \cite{TKR} ) \be\label{R1R2}
   R_1 R_2 =-2\gamma n,
\ee where $n=0,1,2...$.

Substituting next in (\ref{PH1}),(\ref{PH2}), we have
$$
\Phi=\left(%
\begin{array}{c}
  C_1 u_{n-1}(\rho)\\
  iC_2 u_n(\rho)\\
  C_3 u_{n-1}(\rho)\\
  iC_4 u_{n}(\rho)\\
\end{array}%
\right),
$$
and one can fined that coefficients $C_i\,(i=1,2,3,4)$ is
determined by algebraic equations
$$
(E\mp m_1)C_{1,3}-(2\gamma n)^{1/2} C_{4,2} -(p_3 \mp m_2)C_{3,1}
=0;
$$
$$
(E\mp m_1)C_{2,4}-(2\gamma n)^{1/2} C_{3,1}+(p_3 \pm m_2)
C_{4,2}=0.
$$
The equality to zero of the determinant of this system leads to a
spectrum of energy of the non-Hermitian Hamiltonian in the form
\be
    E=\sqrt{{m_1}^2-{m_2}^2 + 2\gamma n +{p_3}^2},
\ee that with take into account $m^2={m_1}^2-{m_2}^2$ coincide
with the eigenvalues of energy of Hermitian Hamiltonian in the
magnetic field \cite{TKR}.

The coefficients $C_i$ may be determined if in as operator of
polarization to choose the component of the tensor polarization in
the direction of the magnetic field \be\label{muH}
          \mu_3=m_1\sigma_3 + \rho_2[\vec{\sigma}\vec{{\cal
          P}}]
\ee where matrices $$ \sigma_3= \left(%
\begin{array}{cc}
  I & 0 \\
  0 & -I \\
\end{array}%
\right); \,\,\,\,\,              \rho_2 = \left(%
                                      \begin{array}{cc}
                                        0 & -iI \\
                                     iI & 0 \\
                                         \end{array}%
                                       \right).
$$

It is easy to see, that bispinor $C$ can be written as

         \be\label{PsiH1} \left(%
\begin{array}{c}
  C_1 \\
  C_2 \\
  C_3 \\
  C_4 \\
\end{array}%
\right)=\frac{1}{2\sqrt{2}}\left(%
\begin{array}{c}
  \cosh(\vartheta/2) \Phi_1+\sinh(\vartheta/2) \Phi_3 \\
  \cosh(\vartheta/2) \Phi_2+\sinh(\vartheta/2) \Phi_4 \\
  \sinh(\vartheta/2) \Phi_1 +\cosh(\vartheta/2) \Phi_3\\
  \sinh(\vartheta/2) \Phi_2 +\cosh(\vartheta/2) \Phi_4, \\
\end{array}%
\right)\ee where
$$
\Phi_1=\sqrt{1+\zeta m/p_\bot}\sin(\pi/4+\lambda/2)
$$
$$
\Phi_2=\zeta\sqrt{1-\zeta m/p_\bot}\sin(\pi/4-\lambda/2)
$$
$$
\Phi_3=\zeta\sqrt{1+\zeta m/p_\bot}\sin(\pi/4-\lambda/2)
$$
$$
\Phi_4=\sqrt{1-\zeta m/p_\bot}\sin(\pi/4+\lambda/2).
$$
Here $\mu_3 \psi =\zeta k\psi$, $k=\sqrt{{p_\bot}^2 + m^2}$ and
$\zeta=\pm 1 $ that is corresponding to the orientation of the
fermion spin: along $(+1)$ or opposite $(-1)$ to the magnetic
field, and parameter $ \lambda$ obey to the condition
$\cos{\lambda}=p_3/E.$

\section{ Non-Hermitian modified Dirac-Schwinger-Pauli  model in the magnetic
field}

In this section, we will touch upon a question of describing the
motion of Dirac particles, if their own magnetic moment is
different from the Bohr magneton. As it was shown by Schwinger
\cite{Sc}, that  the Dirac equation of particles in the external
electromagnetic field $A^{ext}$ taking into account the radiative
corrections may be represented in the form \be\label{A}
\left({\cal P}\gamma -m\right)\Psi(x)-\int{\cal
M}(x,y|A^{ext})\Psi(y)dy=0, \ee where ${\cal M}(x,y|A^{ext})$ is
the mass operator of fermion in external  field. From equation
(\ref{A}) by means of expansion of the mass operator in series
according to  $ eA^{ext}$ with precision not over then linear
field terms  one can obtain the modified equation( see, for
example, \cite{TKR}). This equation preserves the relativistic
covariance and consistent with the phenomenological equation of
Pauli obtained in his early papers.

Now let us consider the model of massive fermions with
$\gamma_5$-extension of mass $m\rightarrow m_1+\gamma_5 m_2$
taking into account the interaction of their charges and anomalous
magnetic moment(AMM) with the electromagnetic field $F_{\mu\nu}$:

\be\label{Delta} \left( {\cal P}_\mu\gamma^\mu
 -
 m_1 -\gamma_5 m_2 -\frac{\Delta\mu}{2}\sigma^{\mu \nu}F_{\mu\nu}\right)\Psi(x)=0,\ee
where $\Delta\mu = (\mu-\mu_0)/\mu_0$ - AMM of fermion,
$\mu_0=|e|/2m$ - the Bohr magneton,
$\sigma^{\mu\nu}=i/2(\gamma^\mu \gamma^\nu-\gamma^\nu
\gamma^\mu)$. Thus phenomenological constant which was introduced
by Pauli $\Delta\mu $,  is part of the equation and gets the
interpretation with the point of view quantum field theory(QFT).

The Hamiltonian form of (\ref{Delta}) in the homogenies magnetic
field is the following \be i\frac{\partial}{\partial t}
\Psi(r,t)=H_{\Delta \mu}\Psi(r,t),\ee where \be\label{Delta1}
H_{\Delta\mu} = \vec{\alpha}\textbf{{\cal P}} + \beta(m_1 +
\gamma_5 m_2) + \Delta\mu\beta(\vec{\sigma}\textbf{H}).\ee Given
the quantum electrodynamics contribution in AMM with accuracy up
to $e^2$ order we have $\Delta\mu=\frac{\alpha}{2\pi}\mu_0 $,
where $\alpha = e^2 =1/137$ - the fine-structure constant and we
still believe that the potential of an external field satisfies to
the free Maxwell equations.

It should be noted that now the operator projection of the fermion
spin at the direction of  its movement - $\overrightarrow{ \sigma}
\textbf{{\cal P }} $ is not commute with the Hamiltonian
(\ref{Delta1}) and hence it is not the integral of motion. The
operator, which is commuting with this Hamiltonian and it is an
integral of motion remains $\mu_3$ (see (\ref{muH})). Subjecting
the wave function $ \psi $ to requirement to be eigenfunction of
the operator polarization (\ref{muH}) and Hamilton operator
(\ref{Delta1})we can obtain: \be\label{Pi}
\mu_3\psi = \zeta k\psi, \,\,\, \mu_3=\left(%
\begin{array}{cccc}
  m_1 & 0 & 0 & {\cal P}_1-i{\cal P}_2 \\
  0 & -m_1 & -{\cal P}_1-i{\cal P}_2 & 0 \\
  0 & -{\cal P}_1+i{\cal P}_2 & m_1 & 0 \\
  {\cal P}_1+i{\cal P}_2 & 0 & 0 & -m_1 \\
\end{array}%
\right), \ee where $\zeta=\pm 1$ are characterized the projection
of fermion spin at the direction of the magnetic field.
 $$H_{\Delta\mu}\psi=E\psi,$$
  \be\label{Hmu} H_{\Delta\mu}=\left(%
\begin{array}{cccc}
  m_1+H\Delta\mu & 0 & {\cal P}_3 -m_2& {\cal P}_1-i{\cal P}_2 \\
  0 & m_1-H\Delta\mu & {\cal P}_1+i{\cal P}_2  & -m_2-{\cal P}_3\\
  m_2+{\cal P}_3 & {\cal P}_1-i{\cal P}_2 & -m_1-H\Delta\mu & 0 \\
  {\cal P}_1+i{\cal P}_2 & m_2-{\cal P}_3 & 0 & H\Delta\mu-m_1 \\
\end{array}%
\right). \ee

\begin{figure}[h]
\vspace{-0.2cm} \smallskip
\includegraphics[angle=0, scale=0.5]{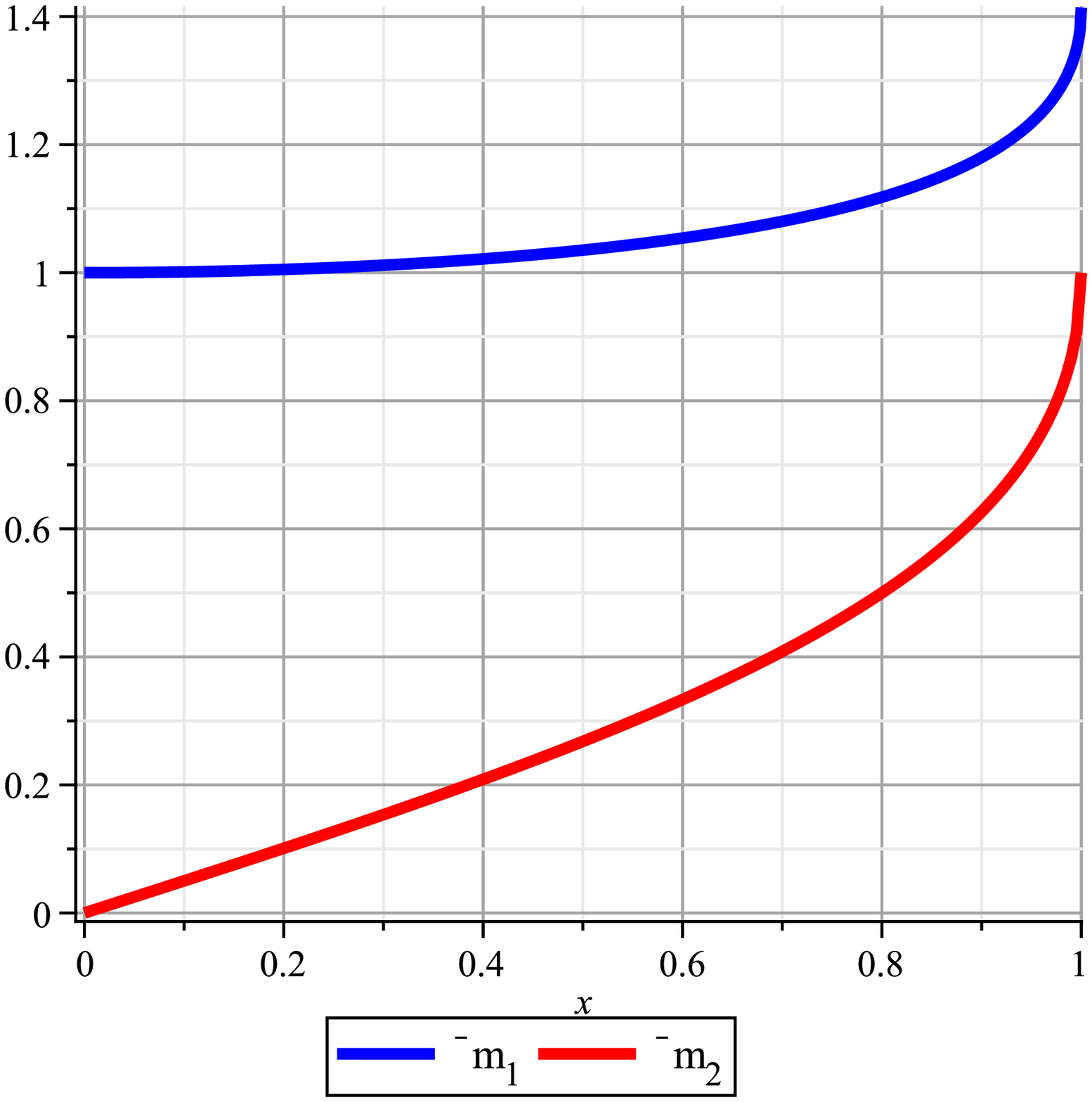}
\caption{The dependence of parameters $^{-}m_1/m, ^{-}m_2/m $ on
the $x=m/M.$} \vspace{-0.1cm}\label{f4}
\end{figure}

\begin{figure}[h]
\vspace{-0.2cm} \smallskip
\includegraphics[angle=0, scale=0.5]{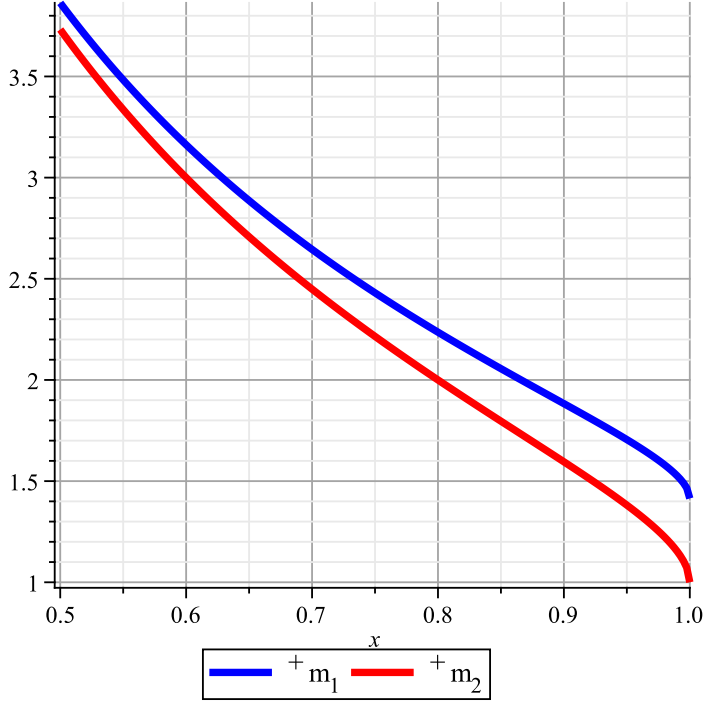}
\caption{The dependence of parameters $^{+}m_1/m, ^{+}m_2/m $ on
the $x=m/M.$} \vspace{-0.1cm}\label{f5}
\end{figure}

 Performing calculations in many ways reminiscent of similar
calculations carried out in the previous section for the fermion
energy may be find the exact solution:

\be\label{E} E(\zeta,p_3,2\gamma
n,H)=\sqrt{{p_3}^2-{m_2}^2+\left[\sqrt{{m_1}^2+2\gamma
n}+\zeta\Delta\mu H \right]^2} \ee and eigenvalues of the operator
polarization we can write in the form \be k=\sqrt{{m_1}^2 +2\gamma
n}. \ee

The  expression analogical to (\ref{E}), in the frame of ordinary
Dirac-Schwinger-Pauli approach was previously  obtained in exact
form in the paper\cite{TBZ}. Direct comparison of modified formula
(\ref{E}) in the Hermitian limit $m_2\rightarrow 0 $ (and
$m_1\rightarrow m$)  with the analogical result \cite{TBZ} shows
their complete coincidence.  It should  also emphasize  that the
expression (\ref{E}) contains dependence on mass parameters $m_1$
and $m_2$, which are not combined into \emph{a mass of particles}
$m$.

Thus, in contrast to the cases described in the previous sections,
here  accounting of interaction AMM of fermions with the magnetic
field allow to raise the question about the possibility of
experimental studies of the influence of Non-Hermitian extensions
of the fermion mass. In particular if to suggest that $m_2=0$ and
hence $m_1 =m$, we return as it was noted early to the Hermitian
limit. But taking into account the expressions (\ref{m11111}) and
(\ref{m22222}) we obtain that the energetic spectrum dependence
(\ref{E}) is expressed through the fermion mass $m$ and the value
of the maximal mass $M$. Taking into account  that the AMM removes
the degeneracy on spin  we can obtain the energy of the ground
fermion state $n=0,p_3 =0,\zeta=-1$ in the form \be\label{E1}
E(-1,0,0,H,x)=m\sqrt{-\left({\frac{1\mp\sqrt{1-x^2}}{x}}\right)^2+
\left(\frac{\sqrt{2}\sqrt{1\mp\sqrt{1-x^2}}}{x}-\frac{\Delta\mu
H}{m} \right)^2}, \ee where $x=m/M$ and the upper sign corresponds
to the ordinary particles and the lower sign defines their
"exotic" partners (see the footnote after formula (\ref{x})).

Through decomposition of functions ${}^{-}m_1$ and ${}^{-}m_2$ we
can obtain

\be\label{sys}{}^{-}m_1/m =\left\{
\begin{array}{c}
  1+\frac{x^2}{8}+\frac{7 x^4}{128},\,x\ll 1 \bigskip\\
  \frac{\sqrt{2}}{x}\qquad \qquad x\rightarrow 1 \\
\end{array}\right.\,\,\,\,\,
{}^{-}m_2/m =\left\{
\begin{array}{c}
   \frac{x}{2}+\frac{x^3}{8}+\frac{x^5}{16},\,\,\,\,\,\,\,\,x\ll 1\bigskip\\
  \frac{1}{x}\qquad\qquad\qquad x\rightarrow 1 \\
\end{array}\right.
\ee

Similarly, for ${}^{+}m_1$ and ${}^{+}m_2$ we have

\be\label{sys1}{}^{+}m_1/m =\left\{
\begin{array}{c}
  \frac{2}{x}-\frac{x}{4}-\frac{5 x^3}{64},\,x\ll 1\bigskip \\
  \frac{\sqrt{2}}{x}\qquad \qquad x\rightarrow 1 \\
\end{array}\right.\,\,\,\,\,
{}^{+}m_2/m =\left\{
\begin{array}{c}
   \frac{2}{x}-\frac{x}{2}-\frac{x^3}{8},\,x\ll 1\bigskip\\
  \frac{1}{x}\qquad\qquad\,\, x\rightarrow 1 \\
\end{array}\right.
\ee

Let us now turn to a more detailed consideration of the fermion
energy in the ground state in the external field. As follows from
(\ref{sys} ) and (\ref{sys1}) function (\ref{E}) not trivial
depends on the parameters $x=m/M$ and $H$. For reasons outlined
above, the effect of magnetic field on the lowest-energy  state of
the fermion with the small mass  $x\ll 1$ (for example, the mass
of ordinary  electron $m$) and also considering the smallness of
the constants of interaction AMM with magnetic field
$\Delta\mu=\alpha/2\pi \mu_0$ we can write

\be\label{E1} E(-1,0,0,H,x)= m\left[
1-\frac{\alpha}{4\pi}\frac{H}{H_c}(1+x^2/8 +x^4 7/128)\right], \ee
where $H_c=m^2/|e|=4.41\cdot10^{13}G$-the critical quantum
electrodynamic field.

 On the other hand, for the case of the "exotic" particles (with a
mass of electron $m$ and $\Delta\mu=\alpha/2\pi \cdot  \mu_0$) in
this limit $x\ll1$ one can obtain a result which significantly
different from previous  one (see (\ref{sys1}) and Fig.3)

\be\label{E2} E(-1,0,0,H,x)= m\left( 1-\frac{\alpha}{2\pi}\frac{
 H}{x H_c}\right). \ee
From (\ref{E2}) should be that the field corrections in this case
is increased substantially. Note that under fixed parameters
intensity of the field and the mass of the fermion from (\ref{E2})
one can judge about of a maximal mass value of $M$.

  We can see from (\ref{sys}),(\ref{sys1}) that the changes of the
parameters ${}^{\mp}m_1 $ and ${}^{\mp}m_2$ occur such a way that
in the point $ x=1$ ($m=M$) functions corresponding to the
ordinary and exotic particles are crossed. At Fig.2 and Fig.3
dependencies ${m}^{\mp}_1/m$ and $m^{\mp}2/m$ on the parameter
$x=m/M$ are represented and one may also clearly see the justice
of this fact.


\section{Summary and
Conclusions.}

 In the researches, presented in the previous
sections, we have shown that the Dirac Hamiltonian of a particle
with $\gamma_5$ -mass term has the unbroken $\cal PT$ - symmetry
in the area ${m_1}^2\geq {m_2}^2$ which however has a number of
sub-regions (see (\ref{ogr})). In this regard, more informative is
the transition from the variables $m_1,\, m_2$  to variables $
m\,\, M$ according to (\ref{m11111}),(\ref{m22222}). In
particular, one is obtained the restriction of  the particle mass
in this model: $m\leq M$. This conditions  describe  the new
boundaries  definition of  the unbroken region $\cal PT$ -
symmetry of  the modified Hamiltonians.

 In addition, we have shown
that the introduction of the postulate about the limitations of
the mass spectrum, lying in the ground of the a geometric approach
to the development of the modified QFT (see, for example
\cite{Max},\cite{KMRS})leads to the appearance of non-Hermitian
$\cal PT $-symmetric Hamiltonians in the fermion sector of the
model with the Maximal Mass. But conversely: using of
non-Hermitian $\cal PT$-symmetric quantum theory with $\gamma_5$
mass term may be considered as conditions of the appearance of the
limitation of the mass particle in a fermion sector of the model.

In particular, this applies to the modified Dirac equation in
which produced the substitution $m\rightarrow m_1+ \gamma_5 m_2 $.
Into force of the ambiguity of the definition of parameters $m_1,
m_2$ the inequality $m_1\geq m_2\geq 0$ describes a particle of
two types. In the first case, it is about ordinary particles, when
 mass parameters  are limited by the terms

\be \label{01}     0\leq m_2 \leq m_1/\sqrt{2}. \ee

In the second area we are dealing with so-called «exotic partners»
of ordinary particles, for which is still accomplished
(\ref{e210}), but one can write

\be \label{02} m_1/\sqrt{2} \leq m_2 \leq m_1. \ee

Intriguing difference between particles of the second type from
traditional fermions is that they are described by the other
modified Dirac equations. So, if in the first case(\ref{01}), the
equations of motion  has a limit transition when $m_2 \rightarrow
0$ that leads to the standard Dirac equation, however in the
inequality (\ref{02}) the such a limit is not there.

Thus, it is shown that the main progress, is obtained by us in the
algebraic way of the construction of the fermion model with
$\gamma_5$-mass term is consists of describing the new energy
scale, which is defined by the parameter $M={m_1}^2/2m_2$. This
value on the scale of the masses and serves as a point of
transition from the ordinary particles to exotic. Furthermore, the
possibility of describing of the exotic particles are turned out
essentially the same as in the model with a maximal mass, which
was investigated by V.G.Kadyshevsky with colleagues on the basis
of geometrical approach.

We have presented a number of examples of non-Hermitian models
with $\gamma_5$-extension mass in relativistic quantum mechanics
including in presence of external electromagnetic field for which
the Hamiltonian $H$ has a real spectrum. Although the energy
spectra of the fermions in some cases were makes them
indistinguishable from the spectrum of corresponding Hermitian
Hamiltonian $H_0$ we found
 example, in which the energy of fermions is clearly dependent
on non-Hermitian characteristics. We are talking about the
consideration of the interaction of the anomalous magnetic moment
of fermions with a magnetic field. In this case we obtained the
exact solution for the energy of fermions (see (\ref{E}).

It should be noted that the formula (\ref{E})  is a valid not only
for charged fermions, but and for the neutral particles possessing
AMM. In this case you must simply replace the square of quantized
transverse momentum of a charged particle in a magnetic field on
the ordinary value $2\gamma n\rightarrow {p_1}^2+{p_2}^2$.
 Thus, for the case of ultra cold polarized exotic electron neutrino we can write
 \be\label{E3}
E(-1,0,0,H,x)= m\left( 1-\frac{\mu_{\nu_e}}{\mu_0}\frac{M
 H}{m_{\nu_e} H_c}\right).
 \ee

It is well known  \cite{n3},\cite{n4} that in the minimally
extended Standard Model  the one-loop radiative correction
generates neutrino magnetic moment which is proportional to the
neutrino mass \be\label{mu1}
  \mu_{\nu}=\frac{3}{8\sqrt{2}\pi^2}|e| G_F
  m_\nu=\left(3\cdot10^{-19}\right)\mu_0\left(\frac{m_\nu}{1
  eV}\right),
\ee where $ G_F$-Fermi coupling constant and $\mu_0$ is Bohr
magneton.
 However, so far, the most stringent laboratory constraints on the
 electron neutrino magnetic moment come from elastic
 neutrino-electron scattering experiments:
$ \mu_{\nu_e} <(1.5\cdot 10^{-10})\mu_0$\cite{n1}.

Besides the discussion of problem of measuring the mass of
neutrinos (either active or sterile) show that for an active
neutrino model we have $\sum m_\nu =0.320 eV$, whereas for a
sterile neutrino $\sum m_\nu =0.06 eV$ \cite{n2}.
 One can also estimate the change of the border of region of unbroken
${\cal P}{\cal T}$-symmetry due to the shift of the lowest-energy
state in the magnetic field, using formula (\ref{E3})
$$
\frac{\mu_{\nu_e}}{\mu_0}\frac{M
 H}{m_{\nu_e} H_c}\leq 1.
$$

 Indeed let us take  the
following parameters of neutrino: the neutrino mass equal to
$m_\nu = 1 eV$ and magnetic moment equal to (\ref{mu1}). If we
assume that the values of mass and magnetic moment of exotic
neutrino identical to parameters of ordinary neutrinos, and also
assuming that the the magnetic fields is equal to $H_c $ we can
obtain the estimate of the border area undisturbed ${\cal P}{\cal
T}$ symmetry in the form

\be\label{E4}
    \frac{M}{m_\nu}\mu_{\nu} =\mu_0.
\ee Omitting the numerical coefficients from
(\ref{E4}),(\ref{mu1}) we can find the magnitude of the limit
value of the parameter $M$: \be\label{E5}
   M=\frac{1}{2 G_F\cdot m}=10^5 \cdot\frac{{m_n}^2}{2
   m}\cong 10^8 GeV,
\ee where $m_n$ - the nucleon mass and $m$ - mass of an electron.

The obtained formulas (\ref{E3}),(\ref{E4}) convincingly shows,
that with their help it is possible, in principle, experimentally
evaluate the possibility of the existence of the Maximal Mass at
low energies. However it should be noted that a considerable
increase of investigated field corrections is connected with
regard to the possible contributions from the so-called exotic
particles, but justice applied calculation methods in these
conditions as a matter of fact require a discussion.

{\bf Acknowledgment:} We are grateful to Prof. V.G.Kadyshevsky for
fruitful and highly useful discussions.

 \end{document}